\documentclass[sigconf]{acmart}

\AtBeginDocument{%
  }

\renewcommand\footnotetextcopyrightpermission[1]{}
\usepackage{algorithm} 
\usepackage{algpseudocode} 
\usepackage{makecell} 
\usepackage{xspace} 
\usepackage[nocomma]{optidef}
\usepackage[T1]{fontenc}
\usepackage{multicol}
\usepackage[normalem]{ulem}
\usepackage{enumitem}
\usepackage{graphicx}
\usepackage{subcaption}
\usepackage{multirow}
\usepackage{lipsum}
\usepackage[all]{nowidow}
\usepackage{listings}
\begin{document}

\title[A Declarative Modelling Framework for the Deployment and Management of Blockchain Applications]{A Declarative Modelling Framework for \texorpdfstring{\\}{} the Deployment and Management of Blockchain Applications}

\author{Luciano Baresi\texorpdfstring{$^*$}{}, Giovanni Quattrocchi\texorpdfstring{$^*$}{}, Damian Andrew Tamburri\texorpdfstring{$^+$}{}, Luca Terracciano\texorpdfstring{$^*$}{}}
\affiliation{%
  \institution{\texorpdfstring{$^*$}{}Dipartimento di Elettronica, Informazione e Bioingegneria, Politecnico di Milano, Milan, Italy, {name.surname}@polimi.it}
  \institution{\texorpdfstring{$^+$}{}Jheronimus Academy of Data Science, Tilburg University, 's-Hertogenbosch, Netherlands, d.a.tamburri@tue.nl}
  \country{}
}

\renewcommand{\shortauthors}{Baresi, Quattrocchi, Tamburri, Terracciano}

\newcommand{\luc}[1]{\textcolor{red}{\textbf{[LB] #1}}}
\newcommand{\gio}[1]{\textcolor{orange}{\textbf{[GQ] #1}}}
\newcommand{\dam}[1]{\textcolor{cyan}{\textbf{[DAT] #1}}}
\newcommand{\lt}[1]{\textcolor{green}{\textbf{[LT] #1}}}

\lstdefinestyle{yaml}{
     basicstyle=\color{blue}\footnotesize\ttfamily\bfseries,
     rulecolor=\color{black}\normalfont\ttfamily,
     string=[s]{"}{"},
     stringstyle=\color{red},
     comment=[l]{:},
     commentstyle=\color{black}\normalfont\ttfamily,
     morecomment=[l]{-},
     numbers=left, numberstyle=\scriptsize,
     frame=single
 }
 
 \lstdefinelanguage{JavaScript}{
  morekeywords={typeof, new, true, false, catch, function, return, null, catch, switch, var, if, in, while, do, else, case, break, const, async, await},
  morecomment=[s]{/*}{*/},
  morecomment=[l]//,
  morestring=[b]",
  morestring=[b]'
}

\lstdefinestyle{htmlcssjs} {%
  basicstyle={\footnotesize\ttfamily},   
  frame=single,
  numbers=left,
  stepnumber=1,
  firstnumber=1,
  numberfirstline=true,	
  identifierstyle=\color{black},
  keywordstyle=\color{blue}\bfseries,
  ndkeywordstyle=\color{black}\bfseries,
  stringstyle=\color{black}\ttfamily,
  commentstyle=\color{brown}\ttfamily,
  language=JavaScript,
  alsodigit={.:;},	
  tabsize=2,
  showtabs=false,
  showspaces=false,
  showstringspaces=false,
  extendedchars=true,
  breaklines=true,
  numberstyle=\scriptsize\color{blue},
  literate=%
  {Ö}{{\"O}}1
  {Ä}{{\"A}}1
  {Ü}{{\"U}}1
  {ß}{{\ss}}1
  {ü}{{\"u}}1
  {ä}{{\"a}}1
  {ö}{{\"o}}1
}

\newcommand{\blue}[1]{\textcolor{blue}{#1}}
\newcommand{\red}[1]{\textcolor{red}{#1}}
\newcommand{\magenta}[1]{\textcolor{magenta}{#1}}
\newcommand{\approach}{\textit{KATENA}\xspace}

\newcommand{\Workflow}{Process\xspace}
\newcommand{\workflow}{process\xspace}
\newcommand{\workflows}{processes\xspace}
\newcommand{\Workflows}{Processes\xspace}

\newcommand{\DeploymentModel}{application model\xspace}
\newcommand{\deploymentmodel}{application model\xspace}
\newcommand{\DeploymentModels}{application models\xspace}
\newcommand{\deploymentmodels}{application models\xspace}

\acmConference[MODELS'22]{ACM Conference}{October 2022}{Montreal,
  Canada}%

\newcommand\blfootnote[1]{%
  \begingroup
  \renewcommand\thefootnote{}\footnote{#1}%
  \addtocounter{footnote}{-1}%
  \endgroup
}

\sloppy
\begin{abstract}


The deployment and management of Blockchain applications require non-trivial efforts given the unique characteristics of their infrastructure (i.e., immutability) and the complexity of the software systems being executed. The operation of Blockchain applications is still based on ad-hoc solutions that are error-prone, difficult to maintain and evolve, and do not manage their interactions with other infrastructures (e.g., a Cloud backend).

This paper proposes \approach, a framework for the deployment and management of Blockchain applications. In particular, it focuses on applications that are compatible with Ethereum, a popular general-purpose Blockchain technology. \approach provides i) a metamodel for defining Blockchain applications, ii) a set of processes to automate the deployment and management of defined models, and iii) an implementation of the approach based on TOSCA, a standard language for Infrastructure-as-Code, and xOpera, a TOSCA-compatible orchestrator. To evaluate the approach, we applied \approach to model and deploy three real-world Blockchain applications, and showed that our solution reduces the amount of code required for their operations up to $82.7\%$.

\end{abstract}



\maketitle

\raggedbottom

\section{introduction}\label{sec:introduction}
\blfootnote{\url{doi.org/10.5281/zenodo.7009710}}
Blockchain has emerged as a new disruptive technology that provides a decentralized and traceable ledger where users can safely record transactions among one another in a trustless environment with high-security guarantees. 
Many Blockchain implementations turned into computing platforms by adding the possibility to develop \textit{smart contracts}: general-purpose computer programs that are stored and executed on the Blockchain. Ethereum~\cite{Buterin2013} is one of the most popular platforms for Blockchain applications~\cite{durieux2020empirical,alharby2018Blockchain} and its runtime, the Ethereum Virtual Machine (EVM), has been adopted by several other Blockchains~\cite{chainlist}.
Recently, many applications, for example, the ones related to finance~\cite{chen2020Blockchain}, supply chain~\cite{helo2019Blockchains}, and collectibles~\cite{ali2021introduction}, have adopted this computation paradigm, and have started to move critical components ---or the whole business logic--- on-chain (i.e., on the Block\-chain).

Smart contracts are interesting but challenging. They are transparent, because the deployed code is publicly accessible and everyone can verify it, are highly available since every node in the network can execute their code, and offer interoperability among applications within the same Blockchain since they share the same infrastructure. In contrast, when a contract is deployed on the Blockchain, its code can no longer be modified. This is critical when bugs are found (e.g., security issues): attackers can easily operate~\cite{DBLP:conf/menacomm/MoubarakFC18} on the bugged contract and there is no way for developers to fix it. To overcome this limitation, \textit{upgradable smart contracts}~\cite{Zheng2021} have been recently presented in the Ethereum ecosystem. They exploit well-known design patterns to allow for the seamless upgrade of smart contracts and are considered a de-facto standard for Ethereum-based Blockchain developments~\cite{proxy}.
Besides the operations required to deploy ``standard'' smart contracts, these patterns also require additional configuration in the deployment scripts, adding a further degree of complexity.

While initially, the majority of Blockchain applications were relatively small, that is, only composed of a few contracts, their complexity is now increasing~\cite{DBLP:journals/spe/WuMHL21}. The deployment of the first applications was ``easy'' and carried out through ad-hoc scripts written in well-known imperative programming languages (e.g., Python, Javascript). Developers had to provide a step-by-step description of how the deployment had to be carried out. These ``simple'' approaches are now unfit to address more complex applications since they are error-prone and lack reusability, portability, and agility in evolving the system.

The growth in the complexity of ``traditional'' applications, and the advent of virtualized resources ---often available through the cloud, has already imposed a paradigm shift for managing application deployment and operation. Imperative approaches have been abandoned in favor of declarative solutions, which only require an \textit{\DeploymentModel}, that is, a specification of the application's architecture and of its deployment. A so-called orchestrator then enacts these specifications to deploy and manage the applications.

The peculiarities of Blockchain-based applications do not allow one to simply reuse these solutions, but there is an emerging need for declarative Blockchain-oriented orchestration solutions~\cite{DBLP:conf/bmsd/HeuvelTDIP21}. 
Cloud solutions assume that the infrastructure is provisioned and managed by users and that applications are ``easily'' upgradable. The Blockchain is different: the infrastructure is abstracted away from users, and applications require a non-trivial engineering effort to be updated. A declarative approach would help both increase the efficiency of Blockchain operations, and provide a unified approach to handle \textit{hybrid} applications, that is, applications whose application logic is executed partially on and partially off the Blockchain.

This paper presents \approach, a declarative framework for the deployment and management of Ethereum and EVM-compatible applications. \approach features a model-driven engineering (MDE) methodology \cite{BrambillaCabotWimmer12} at its core. A metamodel defines a set of  reusable components for the creation of \DeploymentModels, ranging from infrastructural nodes to smart contracts, and other components that run outside the Blockchain (i.e., off-chain). Lifecycle operations (i.e., how to deploy and update a single component) are embedded in each component, to further ease the modeling task.
\approach also provides a set of  \workflows to enact the deployment and management of defined models in a sound and automated way.


To assess \approach, we implemented a prototype based on the standard OASIS TOSCA specification language~\cite{DBLP:books/sp/aws14/BinzBKL14} (Topology and Orchestration Specification for Cloud Applications) and xOpera\footnote{\url{https://github.com/xlab-si/xopera-opera}}, a TOSCA-compliant state-of-the-art orchestrator. 
We then conducted an evaluation based on three case-studies~\cite{1412960991}, that is, three real-world applications coming from different domains and with different architectural complexity. The evaluation targeted explicitly a realistic assessment of the pains and gains of our approach from both a qualitative and a quantitative perspective and compared against state-of-the-art solutions, in particular with development frameworks Truffle\footnote{\url{https://trufflesuite.com/}} and Hardhat\footnote{\url{https://hardhat.org/}}. The comparison reveals that \approach is capable of achieving at least the same results when compared to competing approaches, but with a considerably smaller amount of code. Specifically, as the complexity of featured applications increases, \approach features files with as much as 82.7\% less code compared to existing solutions.

The rest of the paper is organized as follows. Section~\ref{sec:background} introduces the background implied by this work: Blockchain, smart contracts, upgradeable Smart Contracts, and state-of-the-art deployment frameworks. Section~\ref{sec:solution} presents the metamodel, the deployment and management \workflows, and usage scenarios for \approach.
Section~\ref{implementation} describes the prototype and implemented features. Section~\ref{sec:evaluation} discusses the assessment we carried out to evaluate \approach and its prototype. Section~\ref{sec:related-work} surveys some related approaches, and Section~\ref{sec:conclusions} concludes the paper.
\section{background}\label{sec:background}
The following sections introduce the background concepts required to fully understand our solution: Blockchain, smart contracts, and the most significant existing deployment frameworks. The reader familiar with these concepts can easily skip or skim through them.

\subsection{Blockchain}
\label{sec:background:blockchain}

A Blockchain is a decentralized peer-2-peer network of nodes that comprises a ledger of user transactions, organized in an append-only sequence (or chain) of blocks, and a shared state that is the result of all executed transactions.
A transaction represents a set of actions, initiated by a user, that causes a change in the shared state of the Blockchain. Since a Blockchain has a \textit{native currency} (e.g., ether in Ethereum), a transaction can be of a mere monetary nature (e.g., Alice sends 10 ether to Bob), but it can also require computation, by means of smart contracts, whose execution must\,be\,paid\,for.

Blockchain solves the so-called Byzantine Generals Problem \cite{DBLP:books/acm/19/LamportSP19}, that is, the problem of validating a shared state in a trustless decentralized environment. To achieve this, every node maintains a local copy of the Blockchain (i.e., ledger, state, and smart contracts) that is used for validating new transactions. Moreover, Blockchains employ a decentralized consensus mechanism that relies on cryptography and on incentives to reach an agreement among the nodes on which transactions should consider valid. Once the valid transactions are determined, they are packed into a new block and honest participants are rewarded in terms of the native currency. 

To execute a transaction on a Blockchain, a user needs to own a \textit{wallet}, that is, a program that interacts with the Blockchain through transactions. In addition, a wallet must have i) a public key that serves as a public address to locate the wallet and ii) a private key that, in combination with the public one, allows one to create transactions with the wallet. 

Several Blockchain support \textit{smart contracts}, computer programs stored in the Blockchain and executed within it. Some frameworks offer simplified and not Turing-complete programming languages (e.g., Bitcoin Script) while other ones, such as Ethereum and EVM-compatible networks, enable general computation on a Blockchain.

Developers write smart contracts in programming languages developed ad-hoc such as Solidity and Vyper, or in well-known programming languages such as Rust and Javascript. In all these languages, smart contracts have an object-like structure with a constructor, state variables, and functions with their visibility modifiers. Smart contracts can also make use of \textit{libraries}, a particular type of smart contract that has no internal state but only reusable functions that can be called from other contracts. 

Focusing on Ethereum and EVM-compatible blockchains, once developers have finished writing smart contracts and libraries, their code is compiled into \textit{bytecode} that is executable by the EVM. Moreover, a standardized description of the smart contract (or library) interface,  called \textit{Application Binary Interface} (ABI) is generated to allow on-chain and off-chain components to interact with it.

The deployment of a smart contract and the libraries it uses is done by sending a transaction to the network that includes the bytecode. Once the transaction has been accepted, the bytecode and the contract's state are stored in the blockchain at a specific \textit{address}, which allows users and other contracts to locate it. Note that each library is deployed separately from the smart contract at a different address. Smart contracts can also be deleted from the Blockchain, freeing the storage and receiving a reward as an incentive for doing that.

The execution of a smart contract does not come for free: every instruction has a cost (in terms of native currency) that must be paid to discourage flooding the network (and thus trying to prevent Denial of Service attacks). Only users can initiate the execution of contracts by creating a transaction. Within ongoing transactions, then a smart contract can call other smart contracts.

\subsection{Upgradeable Smart Contracts}

When a smart contract is deployed, it is written in the ledger and cannot be changed anymore. For this reason, in the early days, the code of a smart contract was considered to be \textit{immutable}. However, bug-free code is often a dream, and in the past years, there were cases of money loss due to errors. This led to the use of well-known design patterns to create \textit{upgradeable smart contracts}. While this collection includes several patterns, they all derive from two important ones called \textit{proxy} and \textit{diamond}. 

Pattern \textit{proxy}~\cite{gamma1994design} places a proxy in front of a contract and all the requests are received by the proxy and forwarded to the contract. This proxy is nothing but another, simpler, contract with no application logic. When a new version of a contract is deployed, the proxy is configured to forward the requests to it, while users can keep sending requests to the same proxy.

\begin{figure}[t]
    \centering
    \includegraphics[width=1\linewidth]{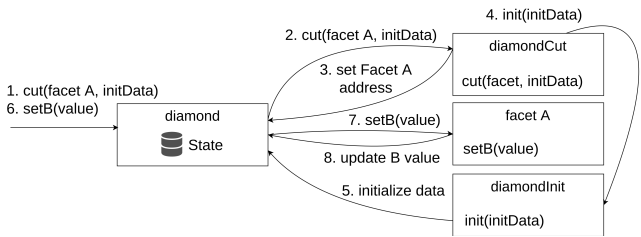}
    \caption{Pattern Diamond.}
    \label{fig:background:diamond-composition}
    \Description{Diamond composition and state upgrade.}
\end{figure}

A pattern that fully exploits this principle is called \textit{diamond}~\cite{Diamond}, also known as \textit{multi-facet proxy}. The core idea is identical to the one used in \textit{proxy} but is extended to enable calls to multiple smart contracts. Developers deploy i) a proxy-like contract, called \textit{diamond}, that contains the application state but no logic, ii) a set of smart contracts, called \textit{facets}, each of them including a part of the application logic but no state, and iii) two additional ``special'' facets called \textit{diamondCut} and \textit{diamondInit}.  The former is used to add and remove facets, while the latter is used by diamondCut to initialize the diamond's state whenever its composition (i.e., its facets) changes. Through a sophisticated delegation mechanism, contract diamond can call all the functions declared in the facets which in turn will alter the diamond's state. 

These concepts are exemplified in Figure \ref{fig:background:diamond-composition}. The diamond contract contains the application state. Users can add a facet ($facet\, A$ in the figure) calling function $cut$ (step $1$ in the figure) that expects the address of a deployed facet and an initial (slice of) state. This call adds the facet to the application by invoking diamondCut (steps $2-3$). diamondCut calls diamondInit that, in turn, updates the state of the diamond (steps $4-5$). Assuming that $facet\, A$ provides a function $setB$ that changes attribute $B$ of the application's state, users can call function $setB$ (step $6$) on the diamond that delegates its execution to $facet A$ (step $7$). Finally, $facet A$ state modifications are reflected on the diamond's state (step $8$).

This pattern allows to separate the application logic in multiple contracts (i.e., facets) that can be added and removed at will. Moreover, the same facets, being stateless, can be reused by multiple diamonds.

\subsection{dApps and their Operations }\label{sec:chainops}

A Blockchain application is often called \textit{decentralized application} or dApp. These applications can be grouped into: \textit{fully decentralized Applications} (f-dApp) and \textit{hybrid decentralized applications} (h-dApp). The former deploys all the application logic  \textit{on-chain} and its only off-chain component is the frontend/client that is usually hosted on decentralized storage (e.g.,  IPFS\footnote{\url{https://ipfs.io}}). The latter uses Blockchain components only partially and a subset of the application logic is deployed off-chain (e.g., on a Cloud backend) along with its frontend.

Many frameworks exist to help developers create dApps: for example, Truffle or Hardhat. Besides overseeing the compilation of smart contracts, they handle their deployment through scripts written using imperative programming languages, like Javascript or Python. These scripts require users to provide a step-by-step description of the deployment. In particular, one must take into account: (i) the links between smart contracts and required libraries, (ii) the deployment order of contracts, and (iii) the connections between contracts that one can call the others when needed through their addresses. For example, to deploy smart contracts $A$ and $B$, where $A$ can call $B$ and $B$ uses library $L$ then the following steps must be executed: i) link library $L$ in $B$, ii) deploy $B$ and retrieve its address, iii) deploy $A$, and iv) store the address of $B$ in $A$. Moreover, when a new version of $B$ is deployed, all the steps except for the third must be (re)executed. This usually requires creating a new script for the upgrade of each component.

Deployment scripts should also set up the connection to an active Blockchain node, i.e., a \textit{Blockchain endpoint}. 
This node can be either \textit{self-hosted} or provided by so-called \textit{node service providers} such as Infura\footnote{\url{https://infura.io}} or Alchemy\footnote{\url{https://www.alchemy.com}}.
Users can host Blockchain nodes on a private server/device using a dedicated node implementation (e.g., Geth\footnote{\url{https://geth.ethereum.org}}). On the one hand, this way they have direct access to the Blockchain network without any intermediary, on the other they must maintain the node which could be complex and expensive (e.g., run security updates, pay for electricity).
Using node service providers is easier but introduces a centralized party and additional information is required to perform operations (e.g. authentication with access keys).

Deployment scripts must take into account these differences, resulting in an additional complexity to be tackled by developers. Moreover, users must take into account additional steps for \textit{h-dApps} including the deployment and configuration of off-chain components and the connection between them and on-chain ones (e.g., setting in a Cloud backend the address of a smart contract that is used during an off-chain computation). None of the existing development frameworks allows to do so and different tools must be integrated to set up the whole system.




\section{\approach}
\label{sec:solution}

\approach is based on a metamodel that defines a set of resuable on- and off-chain components to model dApps: users create \DeploymentModels by instantiating and composing them. Each element defines the characteristics of a component type and the operations needed to manage its lifecycle (e.g., pre- and post-deployment logic).
The dependencies among components are also part of the \DeploymentModel and create a \textit{dependency graph} that constrains the order in which components are managed. We identified 5 different dependency types: \textit{Library-Library} (L-L), \textit{Library-Contract} (L-C), \textit{Contract-Contract} (C-C), \textit{Lazy-Contract-Contract} (Lazy-C-C), and \textit{Off-chain-On-chain} (O-O).



\textbf{Library-related dependencies (L-L, L-C).} When a library is used in a smart contract (or in another library), the library address should be directly embedded in the contract bytecode. Since the address is not known a-priori (i.e., it will only be known after the deployment of the library), its bytecode is filled with placeholders in place of the addresses. After the deployment of the library, developers need to replace placeholders with its address. 

\textbf{Contract-related dependencies (C-C, Lazy-C-C).} Unlike libraries, when a smart contract needs to communicate with another, the first stores the address of the second in a state (or storage) variable.
In C-C dependencies the address is passed to the constructor and the state variable is initialized at contract creation time (i.e., when the constructor is called). Lazy-C-C dependencies cover the case where the association is materialized only when a dedicated function (e.g., a setter) is called. A pair of smart contracts may be connected by both a C-C and Lazy-C-C dependencies, in case the implementation supports both the scenarios.
While C-C dependencies impose a strict deployment order since one contract needs the address of the other at creation time, Lazy-C-C dependencies are more ``dynamic'' and the two contracts may be deployed in parallel. However, only when both are successfully deployed the association can be materialized through a function call.

\textbf{Off-chain-On-chain dependencies (O-O).} Off-chain components need two pieces of information to communicate with a smart contract. The first one is a Blockchain endpoint, which does not create a significant dependency since is known beforehand. The second is the smart contract address, which forces the deployment of the off-chain components to be postponed until the deployment of the smart contracts is completed.

It must be noted that dApp components may use already deployed smart contracts belonging to other applications. We did not describe these kinds of dependencies herein because they do not affect the deployment and operations order. However, our metamodel, described in the following, provides means to model such associations.

\subsection{Metamodel}

\begin{figure*}[t]
    \centering
    \includegraphics[width=0.75\textwidth]{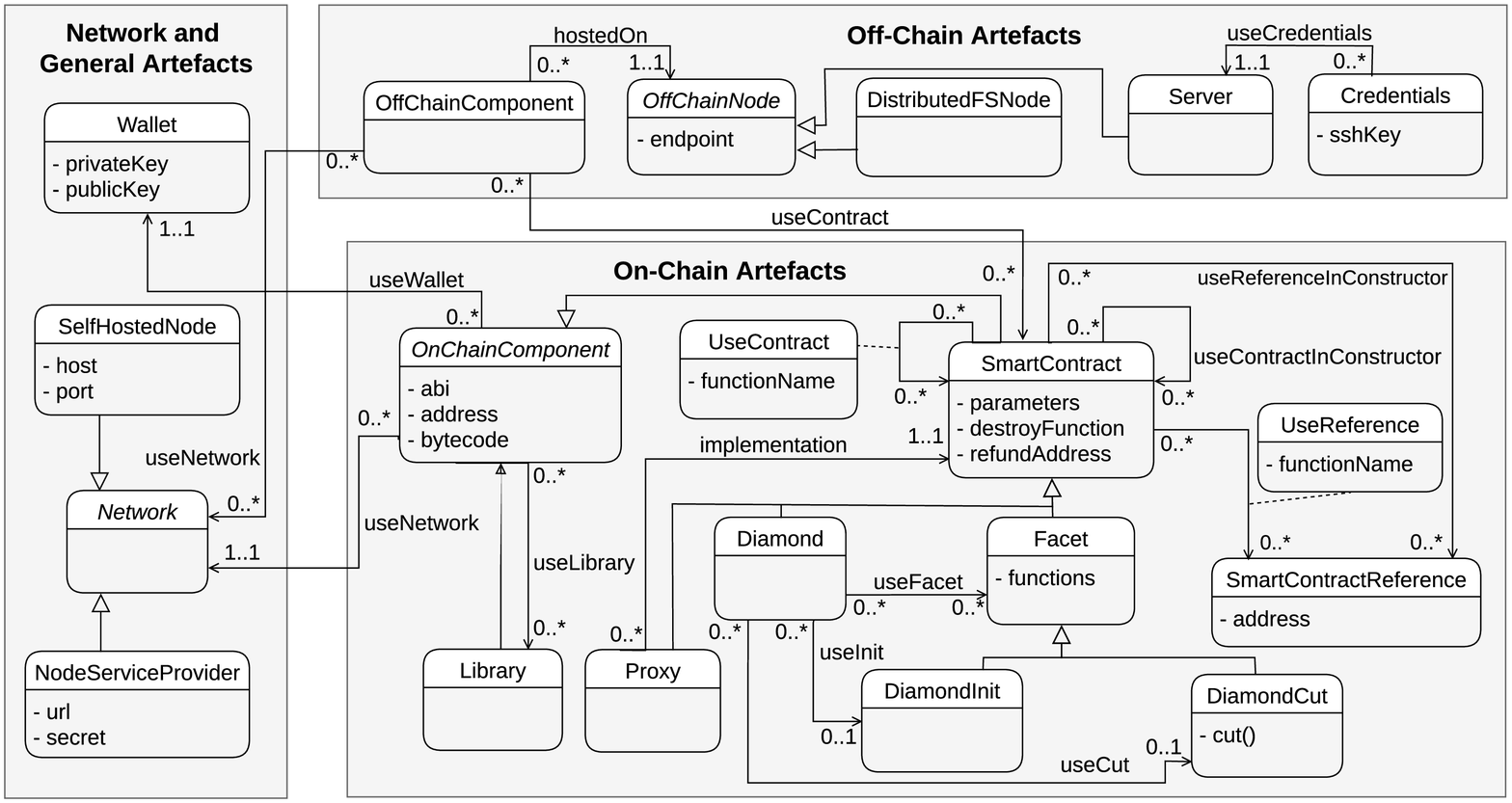}
    \caption{\approach metamodel.}
    \label{fig:solution:metamodel}
    \Description{\approach.}
\end{figure*}

Figure \ref{fig:solution:metamodel} shows the \approach metamodel ---as UML class diagram. 
All attributes are private and their respective getters and setters are omitted for brevity. All the components are assumed to have internal methods (not reported for the sake of brevity) that can handle their lifecycle. They also have constraints between them.
For example, a smart contract \verb#A# cannot require in its constructor the address of smart contract \verb#B# that, in turn, uses \verb#A# in its constructor. Our prototype (Section~\ref{implementation}) implements the lifecycle operations, checks these constraints and aborts the deployment process if they are violated.


The diagram, which features the main architectural elements of f- and h-dApps, is organized around three groups of elements: i) network and general artefacts, ii) on-chain artefacts, and iii) off-chain artefacts.

\textbf{Network and General Artefacts.}  These entities model how applications and users operate on a Blockchain infrastructure.  \textit{Network} is the core (abstract) entity of the group and it is responsible for providing an entry point for Blockchain interactions and operations. Subclasses of \textit{Network} are \textit{SelfHostedNode} and \textit{NodeServiceProvider} that cover the two main setups for accessing a Blockchain network described in Section~\ref{sec:chainops}.
The former provides attributes \textit{host} and \textit{port} that are used to connect  to a specific Blockchain node. This node could be either part of a public or private EVM-compatible network (e.g., an owned Ethereum node),  or a local development environment such as Ganache~\footnote{\url{https://trufflesuite.com/ganache/}}) that provides means to create a local Blockchain for testing activities.
The latter allows interacting with a Blockchain through a node service provider. It requires two attributes: the \textit{url} of the provider, and a \textit{secret} (e.g, an access key) for authenticating the user. 
The user's wallet is modelled as a separate entity (\textit{Wallet}) that stores the \textit{publicKey} and the \textit{privateKey} to access the funds and pay for the transactions required to deploy the smart contracts and manage them (e.g., dependency management).

\textbf{On-chain Artefacts.} These entities model the assets to be deployed on the Blockchain. Their common characteristics are embedded in the abstract class \textit{OnChainComponent}. This class provides attributes and associations that are populated at different times: some are known before deployment, others only at specific steps of the deployment (e.g., after a contract has been deployed).
Attribute \textit{abi} is known beforehand since it stores the interface produced by the smart contract compilation. Attribute \textit{bytecode} is also known, but note that it could be subject to updates during the deployment in case of L-L or L-C  dependencies. We model these two dependencies with association \textit{useLibrary}. 
\textit{OnChainComponent} is deployed and managed on a given Blockchain (association \textit{useNetwork}). Association \textit{useWallet} defines the wallet in charge of operating an instance of \textit{OnChainComponent}.

Entity \textit{SmartContract} is the basic building block to model generic smart contracts. Attribute \textit{parameters} defines the inputs to be passed to the contract constructor. Some of these parameters may be addresses of other smart contracts. \approach models two cases: i) the address belongs to an external contract already deployed on the Blockchain (e.g., an integration with another on-chain application), or ii) a contract of the application to be deployed (Contract-Contract dependency). To model the first case, users must define association \textit{useReferenceInConstructor} with an instance of class \textit{SmartContractReference} that only contains the address of the contract to interface with.
The second case impacts the dependency graph and must be modeled using association \textit{useContractInConstructor} with another instance of \textit{SmartContract}.

Attributes \textit{destroyFunction} and \textit{refundAddress} help specify the function that activates contract destruction and the address to send money (in the native currency), respectively. 

\textit{SmartContract} provides two additional  associations: \textit{useContract} and \textit{UseReference}. The former models Lazy-Contract-Contract dependencies and their management after deployment. The association provides attribute \textit{functionName} to identify the function to invoke to set the required address. Similarly, the latter models the case where the smart contract to be used belongs to an external application whose address was not passed as parameter in the constructor. This association connects the smart contract to an instance of \textit{SmartContractReference}.

Class \textit{Library} inherits from \textit{OnChainComponent} and models libraries and their associations with other components. Libraries do not have a state, a constructor, and they cannot be removed from the Blockchain. However, as described above, they can be connected to other components using association \textit{useLibrary}.

Upgradeable smart contracts are modeled by means of special-purpose entities. \textit{Proxy} provides an abstraction of standard ERC-1967~\cite{proxy}. Besides the properties inherited from \textit{SmartContract}, association \textit{implementation} models the reference to the underlying contract that implements the application logic.

Entity \textit{Diamond} represents the diamond contract of the omonymous pattern. It is connected to instances of facets (entity \textit{Facet}) using association \textit{useFacet} that inherits from \textit{useContract} and materializes a Lazy-Contract-Contract dependency. Attribute  \textit{functions} corresponds to a list of functions that can be called through an instance of \textit{Diamond}. 

Two dedicated classes are created to properly model the diamond pattern: \textit{DiamondCut} and \textit{DiamondInit}. They inherit from \textit{Facet} and they are directly linked to \textit{Diamond} through associations \textit{useCut} and \textit{useInit}, respectively. 

\textbf{Off-chain Artefacts.} Class \textit{OffChainComponent} defines an abstract  off-chain component of the dApp. 
Association \textit{useNetwork}   
connects an instance of \textit{OffChainComponent} to a Blockchain network. An off-chain component can use multiple smart contracts using association  \textit{useContract} that materializes \textit{Off-chain-On-chain} dependencies. 

Components deployed off-chain require the smart contracts' ABI to interact with them. This aspect has not been modelled since  ABIs are usually already required during the development of the off-chain components and so included already in its source code.

An \textit{OffChainComponent} can be hosted on different infrastructures and may require different technologies for its deployment. To offer a unified approach for managing different infrastructures, \textit{OffChainComponent} is linked to an abstract \textit{OffChainNode} by means of association \textit{hostedOn}. Off-chain nodes are also specialized into  \textit{DecentralizedStorage} for f-dApps 
and \textit{Server} for h-dApps. The \textit{Server} node is then linked to \textit{Credential} by means of association \textit{useCredentials} to obtain attribute \textit{sshKey} required to connect to the server.

\subsection{Deployment and Management}
\label{sec:solution:deploy-runtime-management}

\begin{figure}[t]
    \centering
    \includegraphics[width=1\columnwidth]{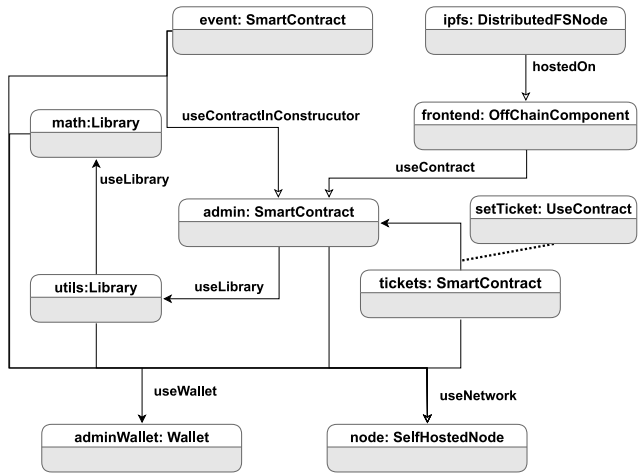}
    \caption{An instance of \approach metamodel.}
    \label{fig:solution:activities}
\end{figure}

The proposed metamodel allows users to create the \DeploymentModel of their dApp through the composition of its elements, but it does not encapsulate the mechanisms to process the entire model. \approach supplies two \workflows to define how to handle the deployment and management of Blockchain applications.
This way, users work at a higher abstraction level since they are not required to define \textit{how} the deployment should carried out. They just need to instantiate the components of interest along with their dependencies.
Figure \ref{fig:solution:activities} shows an object diagram of a dApp that includes all the types of dependencies discussed at the beginning of Section \ref{sec:solution}. It uses a frontend deployed on IPFS, a self-hosted node to connect to the network, and a wallet to perform deployment transactions.

The first \workflow defines the deployment of a Blockchain application.
It starts by setting up the connections with the on- and off-chain infrastructures using subclasses of \textit{Network} and \textit{OffChainNode}, respectively.
Then, it creates the dependency graph by traversing the relevant associations in the metamodel (e.g., \textit{useContractInConstructor}, \textit{useLibrary}).
This graph is then used to oversee the deployment of components, starting from those without dependencies to the ones that use them, recursively, until all the dependencies are satisfied.

First, \approach deploys libraries with no dependencies (e.g., library $math$ in the figure), and their \textit{address} is retrieved.
Then, these addresses are used to solve all the L-L dependencies (e.g., between libraries $utils$ and $math$).

After libraries, \approach deploys smart contracts starting with the ones with no dependencies, followed by the ones with only C-L dependencies (e.g., contract $admin$), and finally, with contract-related dependencies (e.g., contract $events$)  until all on-chain components are deployed. After that, Lazy-C-C dependencies (e.g., between contracts $tickets$ and $admin$) are solved and the setup of the smart contracts is completed. 
Once all smart contracts are deployed, \approach instantiates off-chain components, by exploiting the required information to interact with on-chain components, that is, Blockchain endpoint, and the contract addresses.

The second \workflow is dedicated to managing and upgrading deployed components, activities that may require redeploying existing components.
\approach first identifies the node to be replaced and initiates the deployment of its upgraded version.
If the component is a library, \approach solves L-L and L-C dependencies. In both cases, since these components do not provide ways to change the dependency dynamically, \approach automatically deploys again the dependent libraries and smart contracts by replacing in their bytecode the old address with the one of the upgraded library.
After that (or if the new component is a smart contract), Lazy-C-C and C-C dependencies are managed. In the first case, \approach simply calls the defined function to update the dependency. In the second one, the smart contract must be redeployed so that the address of the upgraded library is properly passed to its constructor.
Once on-chain components have been updated, \approach manages O-O dependencies by setting, where needed, the addresses of the updated or redeployed on-chain components.

\subsection{Usage scenarios}
\label{sec:solution:keys}
We can now envisage three possible usage scenarios: \textit{local device},  \textit{private server}, and \textit{public server}.

\approach can be adopted on a local device (e.g., a personal computer). The keys are held by the application owner (i.e., a single person) and stored on the device. Wallet keys are never sent to private or public servers over the internet and therefore, the risk of malicious attempts to steal them is low. On the other, only one person can deploy elements.

If we think of a private server, \approach can be hosted on a private server, where a dedicated wallet can be created to store company funds and its credentials can be saved in a secure part of the infrastructure. Compared to the case above, this approach enables a permissioned deployment strategy where only people working in the company or automated CI/CD pipelines can deploy the applications.

Finally, the last case is where \approach is offered as-a-Service, that is, it is hosted on a public infrastructure and can be used by many users at the same time.
This scenario presents a trade-off between security and the ownership of the contract. When a new contract is deployed, the Blockchain stores the publicKey of the wallet that initiates the transaction. If the deployment is delegated to a third-party, the creator of the smart contract cannot be recognized as the ``owner'' of the contract. The only way for an external organization to ``impersonate'' another wallet is to have access to its private keys.

This leads to two cases. In the first, users upload their private keys to the public platform and \approach uses them to execute the transactions for the deployment. This way users remain owners of their contracts but they have to trust the platform.
In the second case, the platform has its own wallet where users send money to perform operations. This avoids the disclosing of wallet information but they lost contract ownership. 

An additional solution is to create contracts with dynamic ownership. For example in Ethereum, the standardized   \textit{Ownable}\footnote{\url{https://eips.ethereum.org/EIPS/eip-2767}} contracts allow to change contract ownership with an additional transaction (i.e., by invoking a dedicated function). This way, users can rely on the platform wallet to deploy the contracts without exposing their private keys, and instruct \approach to perform an additional transaction after the deployment is completed that changes the contract ownership to their public address.


\section{Prototype}
\label{implementation}

While the metamodel and \workflows we defined are technology- and language-agnostic, to assess \approach, we exploited TOSCA~\cite{DBLP:books/sp/aws14/BinzBKL14} and xOpera, a TOSCA orchestrator to realize a prototype\footnote{Source code and experiments at: \url{https://zenodo.org/badge/latestdoi/472717537}.}. 

TOSCA is a well-known standard language for modelling cloud applications and infrastructures ---that is, for specifying off-chain components--- in the form of YAML specifications. 
Its main entity type, called \textit{node}, is used to model programming and computational elements, such as executables, middleware infrastructures, and physical or virtual computing resources. A node has \textit{properties} and \textit{attributes}. The former are values provided by users used to create a node instance, while the latter are values that are set by the orchestrator during the deployment or operation of the node.
One can also define some \textit{requirements}, that is, a set of \textit{relationships} that must be satisfied before the node can be created. A relationship is a link between two nodes that allow them to access the other node's properties and attributes.  Relationships are directed edges and they have properties and attributes. 
TOSCA distinguishes between node/relationship types (i.e., blueprints) and instances. Both nodes and relationship types can inherit from already defined types to increase re-usability.
Both nodes and relationships have an associated \textit{interface} that defines a sequence of lifecycle operations. Each operation can be paired with scripts that are executed at proper time through Ansible\footnote{\url{https://www.ansible.com/}}, a well-known engine that automates the configuration and deployment of applications.

We materialized the metamodel and \workflows as a set of reusable TOSCA types and interfaces. Users define \DeploymentModels in the form of TOSCA specifications and no code generation is involved.

The prototype supports any EVM-compatible Blockchain, the creation of ``plain'' smart contracts along with proxies, and ones that use the diamond pattern. Moreover, the prototype can manage all the five types of dependencies described in Section~\ref{sec:solution}. 
We materialized each element of the metamodel as either a node or a relationship type derived from the standard ``root'' types of TOSCA. xOpera supports these types and automates their management.

In particular, while mapping our metamodel, we defined node types for entities\textit{Wallet}, \textit{Network}, \textit{SelfHostedNode}, \textit{SmartContract}, \textit{SmartContractReference}, \textit{Proxy}, \textit{Diamond}, and \textit{DiamondCut}. Each of them was provided with properties and attributes.
Moreover, we implemented lifecycle operations of each node type by deriving them from type \texttt{tosca.interfaces.node.lifecy\-cle.Standard}. This interface provides some standardized phases that can be fully customized according to the application domain and they are supported by xOpera. We linked each phase we support to Ansible scripts that invoke Python programs, which, in turn, interact with the selected EVM-compatible Blockchain through library \texttt{Web3.py}\footnote{\url{https://web3py.readthedocs.io/en/stable/}}.

As an example, node type \texttt{katena.nodes.library} defines properties \texttt{abi} and attributes \texttt{bytecode} (generated by \approach after the compilation of the smart contract) and \texttt{address} (retrieved by \approach after the deployment). This type supports the standard phases: i) \texttt{create} compiles and generates the library bytecode, and ii) \texttt{configure} deploys the library on a selected Blockchain.

\approach also defines a set of relationship types that materialize all the associations defined in its metamodel. Node types define how relationships are used by means of requirements. For example node type  \texttt{katena.nodes.smartcontract} comprises its requirements a set of relationships of type \texttt{katena.relationships.useContract\-InConstructor} to support dependencies of type Contract-Contract.

The lifecycle of a relationship is standardized in TOSCA by type \texttt{tosca.interfaces.relationship.Configure} that we employed in our implementation. As an example, relationship \texttt{katena.rela\-tionships.useFacet} provides a property \textit{exclude}  of type \texttt{list} that defines the function names of a diamond facet that should not be callable by a diamond. Moreover, it implements (using Ansible scripts) lifecycle phases: \texttt{post\_configure\_source} that is activated when the source node (i.e., a \texttt{katena.nodes.smartcontract}) completes phase \texttt{configure}, and \texttt{remove\_target} activated when the target node (i.e., a \texttt{katena.nodes.smartcontract}) is removed from the Blockchain.

The definition and the orchestration of the lifecycle phases of both \approach node and relationship types are the key contributions of the prototype. They allow for i) properly resolving the five types of dependencies captured by our metamodel, ii)  implementing the deployment \workflow described in Section~\ref{sec:solution:deploy-runtime-management}, and iii) reducing users' effort in defining deployment models. In particular, users just need to create instances of \approach node types and connect them with proper relationships. The actual deployment and management operations are hidden in the node/relationship logic and executed ``behind the scenes'' by xOpera.

Off-chain components (e.g., cloud infrastructures) are widely supported in TOSCA and they can be easily mapped with the Off-chain Artefacts entities defined in the metamodel. Moreover, thanks to the relationships we defined, existing TOSCA node types (e.g., \texttt{tosca.nodes.Container.Application\footnote{\url{https://github.com/openstack/tosca-parser/blob/master/toscaparser/elements}}} for containerized applications) can be easily mixed with newly defined on-chain ones. 

\begin{figure}[t]
    \centering
\begin{lstlisting}[style=yaml]
ethereum:
  type: katena.nodes.network.ethereum
userWallet:
  type: katena.nodes.wallet
  properties:
    privateKey: { get_input: UserKeyEthereum }
mathLib:
  type: katena.nodes.library
  requirements:
    - useNetwork: ethereum
    - useWallet: userWallet
  properties:
    abi: "MathImpl"
votingContract:
  type: katena.nodes.smartcontract
  requirements:
    - useNetwork: ethereum
    - useWallet: userWallet
    - useLibrary: mathLib
    - useContractInConstructor: randomGeneratorContract
  properties:
    abi: "Voting"
    parameters:
      - 100
      - 0.1
randomGeneratorContract:
  type: katena.nodes.smartcontract
  requirements:
    - useNetwork: ethereum
    - useWallet: userWallet
    - useLibrary: mathLib
  properties:
    abi: "RandomGenerator"
backend:
  type: tosca.nodes.Container.Application
  requirements:
    - dependency:
        node: votingContract
        relationship: useContract
    ...
\end{lstlisting}
    \caption{An exemplar \approach deployment model.}
    \label{fig:exampleimpl}
\end{figure}

\subsection{Example Scenario}

Figure~\ref{fig:exampleimpl} provides an examplar \approach file. The file defines the deployment model of a simple dApp about on-chain voting that is composed of a library, two smart contracts, and a containerized backend. 
At lines $1$-$2$ node instance \texttt{ethereum} is defined with node type \texttt{katena.nodes.network.ethereum} that implements an entrypoint for the Ethereum Blockchain (a subclass of metamodel's entity \textit{SelfHostedNode}), the network where the smart contracts will be deployed.
The user wallet is defined at lines $3$-$6$. This node instance uses type \texttt{katena.nodes.wallet} (entity \textit{Wallet}) and provides property \texttt{privateKey} which value is read from an external input file (omitted for brevity) using the TOSCA \texttt{get\_input} syntax. 

Lines $7$-$13$ declare the library (type \texttt{katena.nodes.library}. \texttt{mathLib}) that provides property \texttt{abi}, and defines two requirements: i) \texttt{use\-Network} to deploy the application on Ethereum, and ii) \texttt{useWallet} to set \texttt{userWallet} as source of transactions and funds for the deployment.
Lines $14$-$25$ define \texttt{votingContract}, the main smart contract (type \texttt{katena.nodes.smartcontract}) of the application. In addition to \texttt{use\-Network} and \texttt{useWallet}, it provides two requirements: i) \texttt{useLibrary} (type \texttt{katena.relationships.} \texttt{useLibrary}) that implements an L-C dependency with \texttt{mathLib}, and ii) \texttt{useContractIn\-Contructor} that materializes a C-C dependency with contract \texttt{randomGenera\-torContract}. Moreover, contract \texttt{votingContract} defines two properties: \texttt{abi} and \texttt{parameters} for calling its constructor.

Similarly lines $26$-$33$ declare smart contract \texttt{randomGenerator\-Contract} which will be deployed before \texttt{votingContract} given the C-C dependency that connects them.
Finally, lines $34-40$ define the dApp backend (details omitted for brevity). This node instance inherits from type \texttt{tosca.nodes.Container.Application}, an existing TOSCA type.  Given that the backend must interface with contract \texttt{votingContract} (O-O dependency), the node instance defines a dedicated requirement. In this case, the connection is materialized through relationship \texttt{dependency}, a generic association between two TOSCA node instances. TOSCA allows specifying the type of this relationship and target node using sub-fields \texttt{relationship} and \texttt{node} respectively.
\section{Evaluation}\label{sec:evaluation}


To assess the feasibility and benefits of \approach, we modelled and deployed three real-world Blockchain applications covering a sufficiently diverse set of control factors including: size (in terms of source lines of code or \emph{SLOC}), used design patterns, number of smart contracts, complexity (measured as number of contracts dependencies), application domain.  These three applications are: i) \textit{Ethereum Name Service}\footnote{\url{https://github.com/ensdomains/ens}} (ENS), a decentralized DNS system running on Ethereum, ii) \textit{DYDX}\footnote{\url{https://github.com/dydxprotocol/solo}}, a well-known Ethereum platform for decentralized finance, and iii) \emph{Dark Forest}\footnote{\url{https://github.com/darkforest-eth/eth}}, a videogame deployed on Ethereum. Table \ref{table:evaluation:applications} shows control factors mapping onto the observed cases, and shows, for each application, the number of smart contracts and the number of dependencies divided by type. 

On the one hand, the modelling itself of these three applications accounts for the completeness of our modelling notation in addressing the domains covered by the target cases. On the other hand, we performed a comparative analysis to show the differences between \approach and state-of-the-art solutions; said comparative analysis features mixed-methods research~\cite{BorEtAl09}. Note that this evaluation focused on the deployment phase and on on-chain components since i) management scripts were not included in the analyzed application repositories, and ii) off-chain components are not public and we could not find information about them (backend logic and components are usually not public). To perform the aforementioned comparative analysis, we compared \approach against two popular Ethereum development and operation frameworks, namely, \emph{Truffle} and \emph{Hardhat}. These technologies relate directly to the observed cases in this study; specifically, \emph{ENS} and \emph{Dark Forest} implement their deployment logic with Hardhat, while \emph{DYDX} uses Truffle.

\subsection{Metrics and Experimental setup}

The main objective of this evaluation is to compare \approach against state-of-the-art solutions. Our focus is on the ease of use of our approach and the required effort to define operation activities.
Many metrics~\cite{DBLP:journals/jss/PalmaNPT20,DBLP:conf/icse/Campbell18} have been proposed to evaluate the readability and complexity of code but, to the best of our knowledge, none of them is applicable to both declarative and imperative languages. 
Most of the metrics focus on imperative constructs like nesting structures, conditional and error handling statements that are not present, by design, in declarative languages.
Thus, we decided to evaluate the benefits of our approach by measuring the \textit{Number of Tokens} (NoT), that is, the ``words'' in a file. In doing so we included programming language keywords, removed commented lines and lines with log commands, and considered symbol ``\texttt{.}'' as a token separator (e.g., statement \texttt{contract.deploy()} contains two tokens). Using this metric it is possible to estimate the effort practitioners should employ in writing deployment and management scripts \cite{DBLP:journals/jss/PalmaNPT20}, and, thus, to fairly compare \approach against state-of-the-art solutions. 

\begin{table}
\footnotesize
\begin{tabular}{|l|c|c|c|c|c|}
\hline
\textbf{Application} & \textbf{Contracts} & \textbf{L-L} & \textbf{C-L} & \textbf{C-C} & \textbf{Lazy-C-C} \\ \hline
\textit{ENS}                  & 4                  & 0            & 0            & 4            & 0                 \\ \hline
\textit{DYDX}                 & 28                 & 0            & 2            & 21           & 10                \\ \hline
\textit{Dark Forest}      & 16                 & 4            & 13           & 1            & 9                 \\ \hline
\end{tabular}
\caption{Used applications.}
\label{table:evaluation:applications}
\end{table}

We also considered other metrics, such as \textit{Lines of Code} (LoC) or the \textit{number of characters} in a file, but the comparison would have not been balanced for the syntax and structural characteristics of the languages employed.
It is well-known that metric LoC does not represent a significant measure of the size and complexity of a program~\cite{DBLP:conf/esem/BaronW020}. Moreover, \approach files are structured on multiple and short lines, while scripts adopted by the other solutions tend to include long and complex statements on single lines. 
The number of characters is not a fair metric too since it is tightly coupled with the programming language keywords and used naming conventions, which would penalize the competitor solutions.


We performed the deployment of the applications on the Ganache local Blockchain for both \approach  and competitors. For \approach orchestration, we used a local xOpera installation.

\subsection{Evaluation Results and Discussion}

\begin{figure}[t]
    \centering
    \begin{minipage}{1\linewidth}
        \centering\textbf{\approach}
       \begin{lstlisting}[style=yaml]
ensRegistry:
  type: katena.nodes.smartcontract
  requirements:
    - useNetwork: ganache
    - useWallet: userWallet
  properties:
    abi: "ENSRegistry"
publicResolver:
  type: katena.nodes.smartcontract
  requirements:
    - useNetwork: ganache
    - useWallet: userWallet
    - usesContractInConstructor: ensRegistry
  properties:
    abi: "PublicResolver"
    parameters:
      - "0x0000000000000000000000000000000000000000"
    \end{lstlisting}
        
    \centering\textbf{Hardhat}
    \begin{lstlisting}[style=htmlcssjs]
const hre = require("hardhat");
const ethers = hre.ethers;
const AD0 = "0x0000000000000000000000000000000000000000";
async function main() {
    const ENSRegistry = 
        await ethers.getContractFactory("ENSRegistry")
    const PublicResolver = 
        await ethers.getContractFactory("PublicResolver")
    const ens = await ENSRegistry.deploy()
    await ens.deployed()
    const resolver = 
        await PublicResolver.deploy(ens.address, AD0);
    await resolver.deployed()
}
    \end{lstlisting}
    \end{minipage}
    \caption{Deployment of ENS.}
    \label{fig:evaluation:ens}
\end{figure}

The Ethereum Name Service is widely used, with over 465K ENS names registered and 180K of them  active~\cite{DBLP:journals/corr/abs-2104-05185}.
It is composed of four contracts that are used to register the DNS domain along with its owner, and to provide additional features such as subdomains.
Its architecture is quite simple, it provides three Contract-Contract relationships between the core registry, named \textit{ENSRegistry}, and the three others. Figure \ref{fig:evaluation:ens} shows a comparison between Hardhat and \approach on the implementation of one of these relationships (between contract \textit{PublicResolver and \textit{ENSRegistry}}). 

\approach implements the relationship with requirement \textit{callsInConstructor} included in the \textit{publicResolver} TOSCA node allowing the orchestrator to understand the dependencies between the two contracts and deploy them one after the other (and not in parallel).
In Hardhat, this is implemented as a sequence of asynchronous calls (using syntax \texttt{async/await}) and through an imperative paradigm. While in \approach the dependencies are \textit{explicitly declared} easing the understanding of the app structure, in Hardhat this only emerges from the semantic of code and the actual dependencies are ``hidden'' within the complexity of the rest of the code. 

The NoT comparison shows that \approach deployment file is composed of 87 tokens, slightly lower than the 95 of Hardhat. While the quantitative improvement is small, qualitatively it can be observed that \approach provides a higher-level approach to the deployment and management of smart contracts. It abstracts away the complexity of the instructions needed to enact the operations, only requires users to prompt simple inputs, and rely on the orchestrator to automate the whole process. In Hardhat, users must do two tasks at the same time: defining the dependencies among contracts and the instructions needed for the deployment, which  require dealing with complex asynchronous code, error handling, and more.

DYDX is one of the largest exchanges for cryptocurrencies\footnote{\url{http://tiny.cc/forbes-crypto}} and it employs 28 smart contracts with the following dependencies: 2 Contract-Library, 21 Contract-Contract, and 10 Lazy-Contract-Contract dependencies.
Compared to ENS, DYDX uses Truffle for its operation and the code is, generally, much more complex given the higher amount of contracts and dependencies. Notably, \approach showed a significant improvement compared to Truffle in our quantitative evaluation.
Our solution led to a reduction of 39.5\% of NoT with 559 tokens against  923 of the original script. This highlights that as the complexity of operations increases, \approach appears to achieve a larger reduction in the efforts required to write the scripts. Once again, this can be intuitively explained by the more abstract and orchestration-driven approach of \approach that significantly simplifies the deployment and management process.

\begin{figure}[t]
    \centering
    \begin{minipage}{1\linewidth}
         \centering\textbf{\approach}
           \begin{lstlisting}[style=yaml]
diamondCut:
  type: katena.nodes.smartcontract
  ...
diamondLoupe:
  type: katena.nodes.diamond.facet
  ...
diamond:
  type: katena.nodes.diamond
  requirements:
    - useCut: diamondCut
    - useFacet: diamondLoupe
        \end{lstlisting}
        
        \centering\textbf{Hardhat}
            \begin{lstlisting}[style=htmlcssjs]
const diamondCutFacet = await deployDiamondCutFacet();
const loupeFacet = await deployLoupeFacet();
const diamondSpecFacetCuts = [
...changes.getFacetCuts('LoupeFacet', loupeFacet),
];
const diamond = await deployDiamond(
    {
      ownerAddress,
      diamondCutAddress: diamondCutFacet.address,
    }, ...);
const initTx = await diamondCut.diamondCut(
                            diamondSpecFacetCuts,
                            ...);
const initReceipt = await initTx.wait();
if (!initReceipt.status) {
    throw Error(`Diamond cut failed`);
}
        \end{lstlisting}
    \end{minipage}
    \caption{Diamond pattern used for Dark Forest.}
    \label{fig:evaluation:df}
\end{figure}

Lastly, Dark Forest uses Hardhat for its operation, and it is composed of 5 libraries and 11 smart contracts. Two significant characteristics of this application are the usage of the Diamond pattern and Library-Library dependencies.
Figure \ref{fig:evaluation:df} shows the difference in constructing a diamond using the two approaches.
Users of \approach can create diamonds with ad-hoc types (\texttt{katena.nodes.diamond}) and requirements (\texttt{useCut} and \texttt{use\-Facet}) which, behind the scenes, automate the wiring of the dependencies and instruct the orchestrator on the instructions to execute. On the other hand, Hardhat users must deal with complex operations ``manually'', and the resulting code does not allow them to simply understand the structure of the application (as shown in the figure).

The NoT comparison provides objective evidence for this intuition showing a reduction of 82.7\% of tokens (304 vs 1765 tokens) when users employ \approach instead of Hardhat. This  result shows that while state-of-the-art solutions do not scale well with operation complexity (i.e., the code to execute them becomes not only longer but also more complex), \approach allows for writing relatively simple scripts that automate these complex processes. This not only reduces user efforts but also makes deployment and management scripts more readable, more maintainable, and less error-prone.

\subsection{Threats to validity}
\label{sec:threats}

We conducted the experiments using three real-world applications showing the advantages of \approach compared to state-of-the-art solutions. However, we must highlight threats that may constrain the validity of obtained results~\cite{DBLP:books/daglib/p/WohlinHH06}.

\textbf{Internal threats.} The experiments were conducted using Gana\-che as target Blockchain. While Ganache is meant to only be used for development purposes, it implements the behavior of the EVM and therefore we do not expect any significant change in the behavior of \approach when used in production-ready Blockchains. The only practical difference is the cost of transactions (in Ganache transactions are free while on a real Blockchain they can be expensive\footnote{\url{https://tinyurl.com/bored-ape-eth-costs}}). However, it must be noted that \approach generates the same transactions (and same deployments) as competitors, thus cost does not represent a critical factor.

\textbf{External threats.} Our quantitative evaluation is based on the NoT metric. This metric approximates the user effort required to write deployment scripts and can be subject to errors. To the best of our knowledge there is no other metric in the literature that could have allowed us to compare declarative and imperative languages. To mitigate this factor, we analyzed \approach and its competitors with a qualitative approach first, and only used the quantitative measurements to confirm our conjectures.

\section{related work}
\label{sec:related-work}

The automation of operations~\cite{DBLP:journals/ife/WursterBFKLSS20} \cite{DBLP:conf/syscon/BensonPR16} is a well-known problem addressed by academia and industry but its application to the Blockchain is at its early stages. 
Industrial approaches, such as Truffle and Hardhat,  have tackled the problem by requiring users to write imperative ad-hoc scripts tailored for each application.

Van den Heuvel et al~\cite{DBLP:conf/bmsd/HeuvelTDIP21} are the first ones to foresee the need for operations in Blockchain applications, proposing a preliminary architecture of a low-code platform to facilitate the development and deployment of Smart Contracts that focuses on the entire development process. Compared to \approach, they only provide an abstract pipeline for the development and operation of Blockchain applications without any sort of implementation. Moreover, they mostly focus on the development phase (i.e., helping users write the application logic), while \approach focuses only on deployment and management of existing applications.

Knecht et al~\cite{DBLP:conf/aims/KnechtS17} propose a tool that performs automated analysis on Blockchain applications. The approach automatically searches for bugs, evaluates the code quality, and verifies properties using formal logic. We consider this work complementary to \approach and could enhance the automation features provided by our solution.

If we focus on the operation of Blockchain applications, to the best of our knowledge, \approach is the first comprehensive solution that provides: i) a declarative approach to define Blockchain applications by employing a metamodel, ii) dedicated activities to handle application deployment and runtime management, and iii) a usable implementation that extends standard tools.
Other existing approaches focus on the modelling of smart contracts but they do not tackle their operation specifically as \approach.
For example, Jiao et al~\cite{DBLP:conf/fase/JiaoL020} propose a semantic modelling framework for smart contracts aimed to find security vulnerabilities through formal verification. Other solutions~\cite{DBLP:conf/dsn/GaramvolgyiKGK18}\cite{tmdscs} propose model-driven approaches to automate the generation of the source code of the smart contract. 

Several solutions have been developed to support the deployment and evolution of complex software systems with model-driven approaches, covering the deployment and runtime management in a wide range of domains (but not Blockchain applications).
For example, some approaches used existing~\cite{DBLP:journals/ibmrd/BreiterBGMSSS14} specification languages or developed new ones~\cite{DBLP:conf/icsoc/HibaB19}\cite{DBLP:conf/models/BergmayrTNWK14} to model cloud deployments. 
Liu et al~\cite{DBLP:conf/cidr/MaoLMF11} address a specific type of architecture called Cloud Data Architectures, modelling the applications in a data-centric perspective. In this work, cloud resources are modelled as structured data, offering the possibility of configuring and executing operations through a declarative language.

All these approaches use either textual or graphical declarative languages to increase usability and reduce error-proneness. Some of the design choices we made in \approach were inspired by these research efforts.

\section{conclusions and future work}
\label{sec:conclusions}

The paper presents \approach, a declarative framework for the deployment and management of Blockchain applications. It provides a metamodel to represent an application with components deployed both on- and off-chain, facilities to model recurring design patterns, and defines required \workflows. We implemented a prototype of \approach that builds on TOSCA and xOpera. Our evaluation shows that \approach leads to a significant (up to 82.7\%) reduction in code writing compared to state-of-the-art approaches.

Our future work comprises the extension of the metamodel to support more design patterns and components. We will also empirically assess how \approach reduces the required time to write deployment scripts compared to state-of-the-art solutions.

\section{Acknowledgements}
Luca Terracciano has been supported by project ICT4Dev, funded by the Italian Agency for Development Cooperation. Luciano Baresi and Giovanni Quattrocchi have been partially supported by SISMA national research project (MIUR, PRIN 2017, Contract $201752ENYB$).

\bibliographystyle{ACM-Reference-Format}
\bibliography{main}
\end{document}